\documentclass[fleqn,twoside]{article}
\usepackage{espcrc2}
\input epsf

\usepackage{graphicx}
\usepackage[figuresright]{rotating}

\newcommand{\AmS}{{\protect\the\textfont2
  A\kern-.1667em\lower.5ex\hbox{M}\kern-.125emS}}

\title{Status of the ANAIS experiment at Canfranc}

\author{S. Cebri\'{a}n\address{Laboratory of Nuclear and High Energy Physics, University of
Zaragoza, 50009 Zaragoza, Spain}\thanks{Attending speaker:
scebrian@posta.unizar.es}, J. Amar\'{e}\addressmark, J. M.
Carmona\addressmark, E. Garc\'{\i}a\addressmark, I. G.
Irastorza\addressmark\thanks{Present address: CERN, EP Division,
CH-1211 Geneva 23, Switzerland}, G. Luz\'{o}n\addressmark, A.
Morales\addressmark, J. Morales\addressmark, A. Ortiz de
Sol\'{o}rzano\addressmark, J. Puimed\'{o}n\addressmark, M.L.
Sarsa\addressmark, J. A. Villar\addressmark }

\begin{document}

\begin{abstract}
The present status of the ANAIS experiment (Annual Modulation with
NaI's) is shown. ANAIS is intended to use more than 100 kg of
NaI(Tl) in the Canfranc Underground Laboratory (Spain) searching
for seasonal modulation effects in the WIMP signal; in a first
stage, a prototype (one single 10.7 kg crystal) has been developed
in order to obtain the best conditions regarding the energy
threshold and the radioactive background in the low energy region
as well as to check the stability of the environmental conditions.
The first results corresponding to an exposure of 2069.85 kg day
show an average background level of 1.2 counts/(keV~kg~day) from
threshold ($E_{thr} \sim 4$ keV, even using one single
photomultiplier) up to 10 keV.
 \vspace{1pc}
\end{abstract}

% typeset front matter (including abstract)
\maketitle

\section{INTRODUCTION}

According to current robust evidences from supernovae and the CMB
radiation, only $\sim$30\% of the density of the Universe must be
due to matter and the rest should consist of unknown species of
dark energy. In addition, most of the matter must be dark,
non-baryonic and cold. Axions and Weakly Interacting Massive
Particles, the so-called WIMPs, are the leading candidates to this
type of dark matter. The lightest stable particles of
supersymmetric theories, like the neutralino, describe a
particular class of WIMPs.

    Direct searches of WIMPs, which are supposedly
filling the galactic halo, are based on the measurement of the
nuclear recoil induced by the elastic scattering off target nuclei
in a suitable detector \cite{Morales}. This process is rare
(interaction rates range from 10 to $10^{-5}$ c/(kg day)) and has
an energy spectrum which decays almost exponentially from a few
keV; therefore, ultra-low background conditions and very good
energy thresholds are mandatory in the direct detection of WIMPs.
But even in these conditions, the WIMP signal is entangled with
the radioactive background; consequently, to confirm the existence
of these particles it would be necessary to identify a distinctive
signature such as the annual modulation of the counting rates due
to the variation in the relative velocity between the Earth and
the halo, produced by the movement of the Earth around the Sun
\cite{Freese88}. The smallness of this effect ($<10\%$) makes
necessary to accumulate as much statistics as possible having a
long time exposure and a large mass of target nuclei, requirement
which may be fulfilled using NaI. The DAMA collaboration has
reported an annual modulation effect \cite{Bernabei00} which
singles out a region of WIMPs in the parametric space
$\sigma$-$m_{W}$, partially excluded by other experiments (CDMS
\cite{Abusaidi00}, EDELWEISS \cite{Benoit01}, IGEX
\cite{Morales02}) and ZEPLIN \cite{zeplin}.

\section{THE PROTOTYPE OF ANAIS }

    ANAIS (Annual Modulation with NaI's) is a large mass experiment
intended to investigate the annual modulation effect in the signal
of galactic WIMPs \cite{anais}. It will be installed in the
Canfranc Underground Laboratory, located in an old railway tunnel
in the Spanish Pyrenees with an overburden of 2450 m.w.e., using
$\sim$100 kg of NaI(Tl) as an improved scaled-up version of a
previous experience \cite{Sarsa97}. Before setting-up the whole
experiment, a prototype is being developed in Canfranc in an
attempt to obtain the best energy threshold and lowest radioactive
background in the low energy region, as well as to check the
stability of the environmental conditions which influence on the
detector response.

    In the ANAIS prototype, one out of the 14 NaI(Tl) detectors stored
underground since 1988 has been used; it consists of an hexagonal
10.7 kg crystal encapsulated inside 0.5-mm-thick stainless steel
and coupled to a photomultiplier (PMT) through a quartz window.
Some components of the PMT have been removed because of their
radioimpurities. The scintillator has been placed in a shielding
consisting of 10 cm of archaeological lead (of less than 9~mBq/kg
of $^{210}$Pb) followed by 20 cm of low activity lead, a sealed
box in PVC (maintained at overpressure to prevent the intrusion of
radon), 2-mm-thick cadmium sheets, and finally, 40 cm of
polyethylene and tanks of borated water. An active muon-veto made
of plastic scintillators is covering the set-up.

    The data acquisition system, based on standard NIM and CAMAC
electronics, has two diffe\-rent parts following the two output
signals implemented from the PMT; the fast signal is recorded
using a digital oscilloscope (500 time bins, 10 ns/bin) while the
slow signal is routed through a linear amplifier and
analog-to-digital converters controlled by a PC through parallel
interfaces, to register the energy of events up to $\sim 1.7$~MeV.

\subsection{Noise rejection}

\begin{figure}[t]
\centerline{ \fboxrule=0cm
 \fboxsep=0.5cm
  \fbox{
\epsfxsize=7cm
  \epsffile{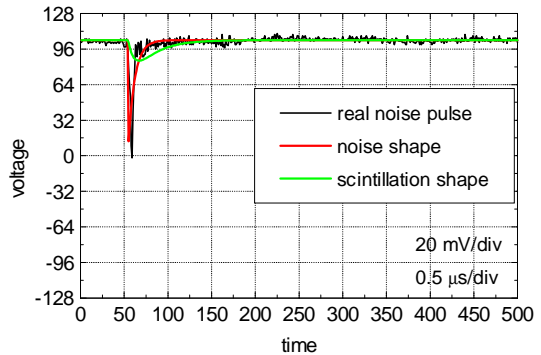}}}
\begin{center}
{\caption {Output pulse from the PMT for a typical noise event and
the theoretical shapes of noise and scintillation for a pulse
having the same area.}\label{pulses}}
\end{center}
\end{figure}

\begin{figure}[t]
\centerline{ \fboxrule=0cm
 \fboxsep=0.5cm
  \fbox{
\epsfxsize=8cm
  \epsffile{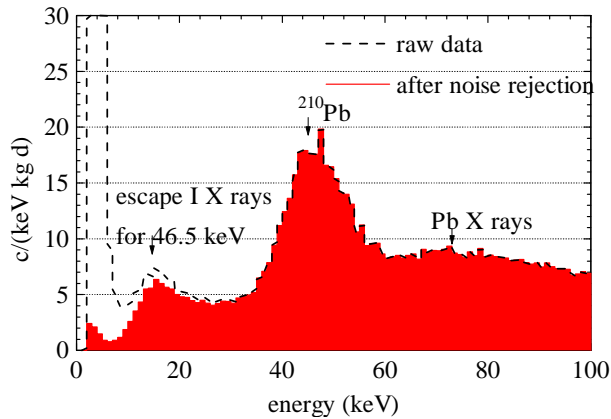}}}
\begin{center}
{\caption {Low-energy region of the observed spectrum in the
prototype of ANAIS before and after the noise
rejection.}\label{spectrum}}
\end{center}
\end{figure}

    The noise in NaI detectors is mainly produced by thermoionic emission
in photomultipliers. The different shape of the output signals
from the PMT for scintillation and noise events makes feasible an
efficient discrimination. Figure \ref{pulses} shows a typical
noise pulse and the theoretical shapes of scintillation and noise
for a pulse having the same area. Typical parameters of the PMT
pulses used in other NaI experiments to reject the noise are the
mean amplitude \cite{Gerbier99}, a ratio of area portions
\cite{Bernabei99}, etc. In the present work, the filtering of
noise uses the squared deviation $d$ of the digitalized pulse
$V_{p}$ from the well-known theo\-re\-ti\-cal shape $V$ of a
scintillation event of the same area:

\begin{equation}
 d=100 \times \frac{\sum_{i}(V(t_{i})-V_{p}(t_{i}))^{2}}{\sum_{i}V_{p}(t_{i})}
 \label{pardc}
\end{equation}
\begin{equation}
    V(t)=-\frac{QR}{\tau-\tau_{s}}(\exp(-t/\tau_{s})-\exp(-t/\tau))
    \label{vt}
\end{equation}
where $\tau=RC$ is the time constant of the RC circuit equivalent
to the PMT, $\tau_{s}$ is the scintillation decay constant
($\sim$230 ns for NaI(Tl)) and $Q$ is the total collected charge.
To eliminate the noise, a safe cut at $3\sigma$ from the center of
the gaussian distribution of this parameter for a population of
$^{137}Cs$ calibration events from 4 to 10~keV has been used. The
effect of the noise rejection can be seen in Figure
\ref{spectrum}, where the raw spectrum and the spectrum after the
elimination of noise are depicted up to 100 keV.

\subsection{Radioactive background}

    By comparing the data recorded from December 2000 to October 2001
with Monte Carlo simulations, it was possible to identify the main
sources of background in the region of interest. The $^{210}$Pb
46.5~keV line as well as a peak due to the escape of X-rays of I
at $\sim$ 16~keV seen in the spectrum (see Figure \ref{spectrum}),
may be caused by the presence of radioimpurities in the stainless
steel can and/or in the PMT. The area of the 1460.8~keV peak is
compatible with an activity of 15~mBq/kg from $^{40}$K in the NaI
crystal; these contaminations produce an almost flat background in
the low energy region due to their beta emission.
% It is also worth noting that comparing the
%spectra observed before and after setting-up the neutron
%shielding, no difference has been noticed; therefore, it seems
%that in this situation the contribution of neutrons to the
%background is negligible.
%, and the main sources come from the detector itself.
It is also worth noting that a comparison between the spectra
recorded with and without the neutron shielding does not show
noticeable differences.

A pulse shape analysis has been carried out in the low energy
region with the purpose of investigating the possible appea\-rance
of the so-called "anomalous" or "bump" events found in other NaI
experiments \cite{Liubarsky00,Gerbier00}.
%Firstly, we have used the method of the UKDMC, and we have
%calculated the time constant of the integrated pulse by fitting it
%to this function. Here we have the distributions of this parameter
%for a population of pulses obtained using a Cs137 calibration
%source and for background events, in the region from 20 to 30 keV.
%For the same pulses, we have obtained also the distribution of a
%different parameter, also used when applying PSD techniques, the
%first momentum of the pulse. As you can see, there is no clear
No evidence of such anomaly has been discovered in the
distributions for background events, neither following the method
of the UKDMC (fitting integrated pulses to calculate the decay
time cons\-tant) nor using other parameters (like the first
momemtum of the pulse).
%are produced by two different
%types of events. The analysis has been performed in other energy
%intervals, obtaining analog results. So, we can conclude that
%there is no evidence of anomalous events in our crystals.

\subsection{Stability}

    In an experiment searching for a very small seasonal modulation,
the stability of the conditions whose variation could mimic the
effect we are looking for is essential. A monitoring system will
be installed in ANAIS to control periodically, and even modify
automatically, the environmental conditions in the laboratory. But
using the data of the prototype, collected along almost 7000
hours, the stability of some parameters has been already checked.

\begin{figure}[t]
\centerline{ \fboxrule=0cm
 \fboxsep=0.5cm
  \fbox{
\epsfxsize=7.5cm
  \epsffile{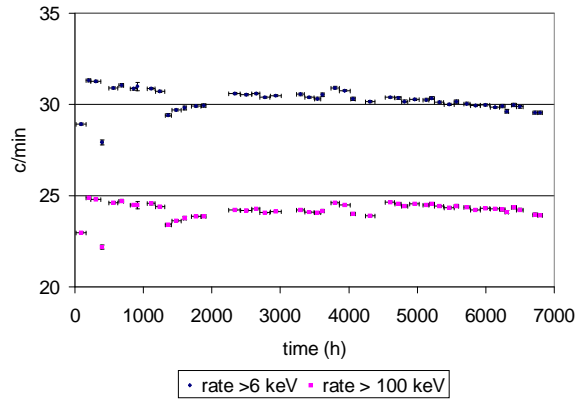}}}
\begin{center}
{\caption {Evolution in time for the whole data collection of the
counting rates integrated from 6 keV and from 100
keV.}\label{rates}}
\end{center}
\end{figure}

    The fluctuations of the ADC channels for peaks used to calibrate
energy spectra are $\sim$1-1.5\%, so the variations in the energy
of events due to this effect are negligible compared to the energy
resolution. The evolution in time of different counting rates
(total, above 6 keV and above 100 keV) has been checked, since a
modulation effect in the low energy region due to background (not
to WIMPs) should be correlated with a modulation at higher
energies. These rates are plotted as a function of time for the
whole data collection in Figure \ref{rates}. The gaussian
distributions of the deviations of the rates from their mean
values have a sigma of 1.75 for the rate above 6 keV, and 1.58 for
the rate above 100 keV.
%, (which are comparable to that obtained by the DAMA experiment.)

    Since the scintillation light yield depends on temperature, a
monitoring system (based on sensors and ADAM sensor-to-computer
interfaces) has been tentatively developed to study the
temperature in the inner enceinte. Internal temperature does not
depend neither on changes on the nitrogen gas flux injected inside
the shielding nor on the periodical fluctuations of the
temperature in the laboratory due to the air conditioning system.
It is only sensitive to real changes outside the shielding. If a
good stability of the external temperature is achieved, variations
inside will be of only 0.03 Celsius degrees.

\section{RESULTS}

The results presented here correspond to an exposure of 2069.85
kg$\times$day. As pointed out before, Figure \ref{spectrum} shows
the raw spectrum and the spectrum after the noise rejection up to
100 keV.
%In detail, we
%have the very low energy region after the noise rejection; an
%estimate of the spectrum after the rejection of electronic recoils
%(assuming the PSD factors obtained on the average) is also
%pictured.
% And on the right, there's a comparison
%of the low energy region of the spectra corresponding to different
%experiments using NaI detectors. In our case,
The energy threshold is of $\sim 4$ keV and the background level
registered from the threshold up to 10~keV is about 1.2
counts/(keV~kg~day).

\begin{figure}[htb]
\centerline{ \fboxrule=0cm
 \fboxsep=0cm
  \fbox{
\epsfxsize=7cm
  \epsffile{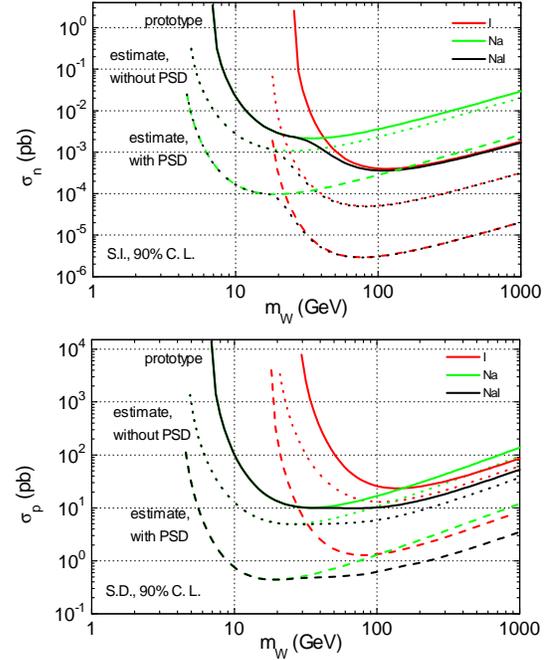}}}
\begin{center}
{\caption {Exclusion plots derived for SI (top) and SD (bottom)
interactions from the prototype (solid line) and expected for the
whole experiment with (dashed line) and without (dotted line) PSD
techniques.}\label{plot}}
\end{center}
\end{figure}

We have used this region to derive the corresponding limits for
the WIMP-nucleon cross sections. The galactic halo is supposed to
be isotropic, isothermal and non-rotating, assuming a density of
$\rho$=0.3 GeV/cm$^{3}$, a Maxwellian velocity distribution with
$v_{rms}$=270 km/s (with an upper cut corresponding to an escape
ve\-lo\-ci\-ty of 650 km/s) and a relative Earth-halo velocity of
$v_{r}$=230 km/s. The Helm parameterization \cite{Engel91} is used
for the coherent form factor, while the approximation from
\cite{Lewin96} is con\-si\-dered for the SD case. Spin factors
($\lambda_{p}J(J+1)$) 0.089 and 0.126 are assumed for Na and I
respectively. Fig. \ref{plot} shows, in addition to the limits
derived from the prototype results (solid lines), the estimates
considering a flat background of 1 count/(keV~kg~day) from 2 to 8
keV after an exposure of 107 kg$\times$y both for raw data (dotted
lines) and assuming Pulse Shape Discrimination (PSD) techniques
(dashed lines); rejection factors obtained on the average by other
groups have been used \cite{PSD}. The plots show the contour lines
for each nucleus, Na and I, as well as the NaI case. That is shown
both for SI scalar interactions (top) and SD WIMP-proton
interactions (bottom). It should be noted that for SI interactions
and using PSD techniques, ANAIS will be able to explore the region
of WIMPs singled out by the possible annual modulation effect
reported by the DAMA colla\-bo\-ra\-tion \cite{Bernabei00}, even
though it is not designed to be an exclusion experiment.
%It has been calculated
%using the traditional method, and the spin factors from the list.
%But this plots below show the limits on proton and neutron
%cross-sections, according to the method proposed in this
%reference, which does not depend on the particular WIMP model
%assummed.

According to these first results, the next steps in the
development of the prototype of ANAIS, currently underway, are the
removal of the present PMT and the steel can so as to install,
instead, two ultra-low background PMT and a 1-cm-thick teflon
enclosure filled with special mineral oil, as in the NAIAD
experiment \cite{Spooner00}. A program of measurements to select
high radiopurity materials has been carried out in Canfranc, using
an ultra-low background Ge detector. The goals of these
modifications are to reduce, as much as possible, the various
sources of background, to diminish the noise by using
anticoincidence read-out (lowering so also the energy threshold)
and to improve the collection of the scintillation light.
%, to achieve a better detection efficiency in the very low energy region.

\section{CONCLUSIONS}

Summarizing, a prototype of the ANAIS experiment has been
installed in the Canfranc Underground Laboratory. The noise
rejection has lead to a $\sim 4$ keV threshold, even using a
single PMT. The background in the low energy region is produced by
$^{210}Pb$ in the detector enclosure and/or the PMT and $^{40}K$
in the NaI(Tl) crystal. The stability is being controlled and new
improvements are being developed before setting-up the large-mass
experiment.

\section*{ACKNOWLEDGEMENTS}
The Canfranc Astroparticle Underground La\-bo\-ra\-to\-ry is
operated by the University of Zaragoza under contract No.
FPA2001-2437. This research was funded by the Spanish Commission
for Scien\-ce and Technology (CICYT) and the Government of Arag\'{o}n.

\end{document}